\documentclass[onecolumn]{IEEEtran}

\usepackage{setspace}
\usepackage{mathptmx}
\usepackage{amsmath}
\usepackage{amssymb}
\usepackage{epsfig}
\usepackage{amsthm}
\usepackage{amsfonts}
\usepackage{subfigure}
\usepackage{stmaryrd}
\usepackage[dvips]{color}
\usepackage{graphicx}
\usepackage{mathcomSTEP}

\newtheorem{definition}{Definition}

\theoremstyle{remark}
\newtheorem{rem}[thm]{Remark}

\allowdisplaybreaks
\begin{document}

\title{
On the Dirty Paper Channel with Fast Fading Dirt
}

\author{%
\authorblockN{%
Stefano~Rini\authorrefmark{1} and Shlomo~Shamai~(Shitz)\authorrefmark{2} \\
}

\authorblockA{%
\authorrefmark{1}
National Chiao-Tung University, Hsinchu, Taiwan,
E-mail: \texttt{stefano@nctu.edu.tw} }

\authorblockA{%
\authorrefmark{2}
Technion-Israel Institute of Technology,  Haifa, Israel,
E-mail: \texttt{sshlomo@ee.technion.ac.il} }
\thanks{
The work of S. Rini was partially funded by the  Ministry Of Science and Technology (MOST) under grant 103-2218-E-009-014-MY2.
The work of S. Shamai was supported by the Israel Science Foundation (ISF) and by the European FP7 NEWCOM\#.
}
}

\maketitle
 \pagenumbering{gobble}
\begin{abstract}
Costa's ``writing on dirty paper'' result establishes that full state pre-cancellation can be attained in the Gel'fand-Pinsker problem
with additive state and additive white Gaussian noise.
This result holds under the assumptions that full channel knowledge is available at both the transmitter and the receiver.
In this work we consider the scenario in which the state is multiplied by an ergodic fading process which is not known at the encoder.
We study both the case in which the receiver has knowledge of the fading and the case in which it does not: for both models we derive inner and outer bounds to capacity
and determine the distance between the two bounds when possible.
For the channel without fading knowledge at either the transmitter or the receiver, the gap between inner and outer bounds is finite for a class of fading distributions which
includes a number of canonical fading models.
In the capacity approaching strategy for this class, the transmitter performs Costa's pre-coding against the mean value of the fading times the state while
the receiver treats the remaining signal as noise.
For the case in which only the receiver has knowledge of the fading, we determine a finite gap between inner and outer bounds for two classes of  discrete fading distribution.
The first class of distributions is the one in which there exists a probability mass larger than one half while the second class is the one in which the fading is
uniformly distributed over values that are exponentially spaced apart.
Unfortunately, the capacity in the case of a continuous fading distribution remains very hard to characterize.
\end{abstract}


\section{Introduction}

In the Gel'fand-Pinsker (GP) model \cite{GelFandPinskerClassic} the output of a point-to-point memoryless channel is obtained as a function of the channel input, a noise term and
a state variable which is non-causally provided to the transmitter but is unknown at the receiver.
In this channel the state may represent the interference caused by another user in a wireless network which is also communicated to the transmitter by the network infrastructure.
In the original setup, both transmitter and receiver are assumed to have perfect channel knowledge: while it is reasonable to assume that a transmitter knows
the channel toward its intended receiver and vice-versa, it is not always realistic to suppose that a transmitter knows the channel between an interfering user and the receiver.
This is especially true in wireless network, since here channel conditions vary continuously over time and reliable channel estimates are hard to obtain.
%
%

The ``writing on dirty paper'' result from Costa \cite{costa1983writing} establishes a closed-form characterization of the capacity of the GP problem in the additive state
 and additive white Gaussian noise setting.
Perhaps surprisingly, the presence of the state does not reduce the capacity of this model, regardless of the distribution or power of this sequence.
In this work we are interested in characterizing  the effect of fading on the capacity of this model and determine the optimal transmission strategies in this scenario.
In the literature, different variations of Costa's setup which also include fading have been considered.
The  ``writing on fading dirt'' channel  in \cite{zhang2007writing} is a variation of the channel of \cite{costa1983writing} in which both the channel input and the state sequence are multiplied by a fading value known at the receiver but not at the transmitter.
The authors of \cite{zhang2007writing} evaluate the achievable region with Costa's assignment and show that the rate loss from full state pre-cancellation is vanishing in both the ergodic and quasi-static fading case.
In the ``compound dirty-paper'' channel of \cite{khina2010robustness} only the state is multiplied by a quasi-static fading coefficient know at the receiver but unknown at the transmitter.
For this model, an inner bound based on lattice strategies is derived to compensate for the channel uncertainty at the transmitter.
Achievable rates under Gaussian signaling and lattice strategies for this channel are derived in \cite{avner2010dirty} while
%
outer and inner bounds to the capacity of the writing on fading dirt channel with phase fading are derived in \cite{grover2007need}.
The approximate capacity of this channel is obtained in \cite{rini2014impact} for the case of binomial and uniform phase fading case.

In this paper we study the ``writing on fading dirt'' model, a variation of the classic model in which the state sequence is multiplied by an ergodic fading coefficient which is not known at the transmitter.
We derive inner and outer bounds to capacity for both the case in which the fading is known at the receiver and for the case in which it is not.
When neither the transmitter nor receiver have fading knowledge, we show that the outer bound can be attained to within a finite gap for a class of fading distribution which
includes the Gaussian, the uniform and the Rayleigh distribution but does not include the log-normal distribution.
%
%
For the case in which only the receiver has fading knowledge, we show a finite gap between inner and outer bound for two classes of discrete distributions:
when the fading distribution has a mass function greater than a half and when it is uniformly distributed over a set of points that are exponentially spaced apart.
%
%
%

The remainder of the paper is organized as follows:
Sec. \ref{sec:Dirty Paper Channel with Phase Fading} introduces the channel model and the some related results.
Sec. \ref{sec:On the capacity of the dirty paper channel with fading dirt and no receiver side information} investigates the capacity for the case in which neither the transmitter nor the receiver have fading knowledge while  Sec.  \ref{sec:Dirty Paper Channel with Fading Dirt and Receiver Side information} focuses on the case in which only the receiver has fading knowledge.
Finally, Sec. \ref{sec:Conclusion} concludes the paper.
%
%
%
\section{Dirty Paper Channel with Fading Dirt}
\label{sec:Dirty Paper Channel with Phase Fading}
\begin{figure}
\centering
\includegraphics[width=.5 \textwidth]{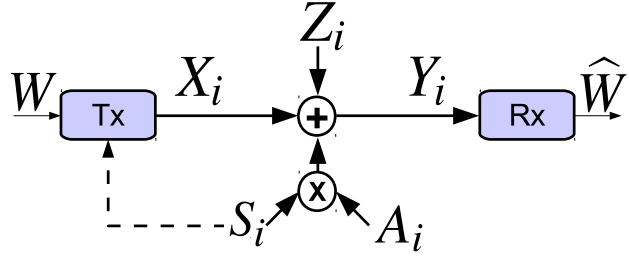}
\vspace{-3.8 cm}
\caption{The Dirty Paper Channel with Fast  Fading Dirty (DPC-FFD).
The dotted line represent the state information provided at the transmitter.
}
\vspace{-.3 cm}
\label{fig:WritingOnFadingDirt}
\end{figure}
In Dirty Paper Channel with Fast Fading Dirt (DPC-FFD), also depicted in Fig. \ref{fig:WritingOnFadingDirt}, the channel output is obtained as
\ea{
Y_i = X_i + c A_i S_i + Z_i, \quad i \in [1 \ldots N ],
\label{eq:WritingOnFadingDirtChannelModel}
}
for $c\in \Rbb$ and where $X_i$ is the channel input, $S_i$ the state,  $A_i$ the fading realization and $Z_i$ the additive noise.
The channel input $X_i$ is subject to a second moment constraint $\Ebb \lsb |X_j|^2 \rsb  \leq P$ while the state $S_i$ and the noise term $Z_i$ are distributed as
\ea{
S_i \sim \Ncal(\mu_S,1), \quad Z_i \sim \Ncal(0,1),  \quad i.i.d.
}
where $\Ncal(\mu,\sgs)$  indicates the Gaussian Random Variable (RV) with mean $\mu$ and variance $\sgs$.
The fading RV $A_i$ is drawn from a distribution $p_A$ which has variance one and mean $\mu_A$.
The state sequence $S^N$ is assumed to be non-causally available at the transmitter while fading sequence $A^N$ is unknown at both the transmitter or the receiver.

\begin{figure}
\centering
\includegraphics[width=.5 \textwidth]{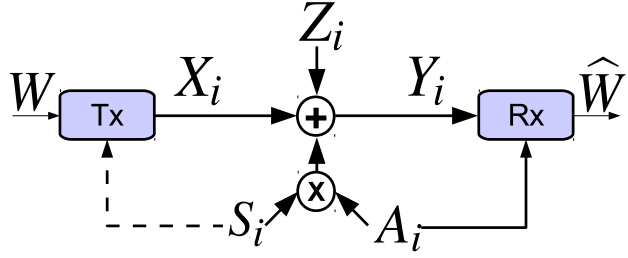}
\vspace{-3.8 cm}
\caption{The Dirty Paper Channel with Fast Fading Dirty and Receiver Channel Side Information (DPC-FFD-RCSI).
}
\vspace{-.3 cm}
\label{fig:WritingOnFadingDirtRCSI}
\end{figure}

A related model to the DPC-FFD in Fig. \ref{fig:WritingOnFadingDirt} is the model in which the fading sequence is provided to the receiver.
%
We refer to this model as the Dirty Paper Channel with Fast Fading Dirty and Receiver Channel Side Information (DPC-FFD-RCSI), also depicted in Fig. \ref{fig:WritingOnFadingDirtRCSI}.
For the DPC-FFD-RCSI the receiver side information can be seen as an additional channel output, that is,
the channel output is the vector $[Y_i \ A_i]$ for $Y_i$ in \eqref{eq:WritingOnFadingDirtChannelModelGeneral}.

%

\begin{rem}{\bf Mean of the state and the fading.}
\label{rem:On the mean of the state and the fading distribution}
The channel output in \eqref{eq:WritingOnFadingDirtChannelModel} can be rewritten as
\ea{
Y
& =X+c \lb  A_0- \mu_A \rb \lb S_0-\mu_S \rb +Z \nonumber \\
& =X + c \lb A_0 S_0 - \mu_A S_0- \mu_S  A_0+ \mu_A \mu_S  \rb +Z,
\label{eq:expression means}
}
where  $A_0= A-\mu_A$ and $S_0= S-\mu_S$. that
%
Each of the term in \eqref{eq:expression means} can be seen as follows

\noindent
$\bullet$ $c \mu_S  A_0 $ can be cancelled at the receiver when it posses fading knowledge.
Without receiver fading knowledge, this term is unknown at both the receiver and the transmitter and is equivalent to additive noise.
%

\noindent
$\bullet$   $c \mu_A S_0$ can be pre-cancelled with Costa coding by the transmitter as in \cite{costa1983writing} (Costa pre-coding in the following).
%

\noindent
$\bullet$ $c A_0 S_0$  requires the cooperation of  both transmitter and receiver, since they each have a knowledge of one of the terms in the multiplication.
%
\end{rem}

The DPC-FFD and the DPC can be used to model the downlink scenario in which a base station is aware of the signal transmitted  by a neighbouring base station but has only partial or no knowledge on the channel between the interference and the intended receiver.
In this scenario it is not clear whether the knowledge of the interfering message is at all useful at the base station since the pre-coding operations heavily rely on the knowledge of the channel gains.

\subsection{Related Results}
\label{sec:Related Results}
\noindent
\underline{\bf Gelfand-Pinsker (GP) channel.}
The DPC-FFD and the DPC-FFD-RCSI are a special case of the GP problem for which capacity is obtained in  \cite{GelFandPinskerClassic}.
\begin{thm}{\bf Capacity of the DPC-FFD/-RCSI \cite{GelFandPinskerClassic}.}
\label{th:Capacity of the ergodic DPC-PF}
%
The capacity $\Ccal$ of the DPC-FFD in \eqref{eq:WritingOnFadingDirtChannelModelGeneral}  is
\ea{
\Ccal=\max_{P_{U,X|S}} \ I(Y; U) - I(U;S),
\label{eq: capacityWritingOnFadingDirt QS}
}
while the capacity of the DPC-FFD-RCSI is obtained from \eqref{eq: capacityWritingOnFadingDirt QS} by considering the channel output $[Y \ A]$.
\end{thm}
The expression in \eqref{eq: capacityWritingOnFadingDirt QS} contains an auxiliary RV $U$ and entails the maximization over the distribution $P_{U,X|S}$. 
%
For this reason a closed-form expression cannot be evaluated easily, either analytically or numerically.
%

\noindent
\underline{{\bf Dirty paper channel with receiver side information and }}
\underline{\bf{phase fading}.}
In \cite{rini2014impact}, we have derived the approximate capacity of the DPC-FFD-RCSI for the case in which $p_A$ is a circularly binomial distribution.

\begin{thm}{\bf Capacity of the DPC-FFD-RCSI with circularly binomial fading \cite[Th. IV.5]{rini2014impact}. \\}
\label{thm:outer bound QS DPC-PF optimized 2 values}
Consider the DPC-FFD-RCSI
\ea{
Y_i = X_i + e^{j \theta_i} S_{R,i} + Z_i, \quad i \in [1\ldots N],
\label{eq:WritingOnFadingDirtChannelModelGeneral}
}
where the state $S_{R,i}$ is a Gaussian RV with zero mean and variance $Q$ and while the fading is $A=\exp \{\theta\}$ for
\ea{
P_{\theta}(t)=\f 12 \lb 1_{\{t=+\Delta\}}(t) + 1_{\{t=-\Delta\}}(t)\rb, \quad \Delta \in [0, \pi/2],
}
then, if $\pi / 4 \leq \Delta \leq \pi/2$, the capacity lies to within  constant gap of  3 bits per channel use from the outer bound
\ea{
& R^{\rm OUT}  =
& \lcb \p{
 \log(P+1)+2
 &  c^2 \leq 1 \\
\f 34 \log(P+1)+2
& c^2 \geq P+1 \\
\f 12 \log(P+1) + \f 12 \log\lb 1+(\sqrt{P}+c)^2  \rb & \\
\  - \f 14 \log(2 c^2) +2
& 1 < c^2 < P+1  \\
} \rnone \nonumber
}
where $c=\sin(\Delta)\sqrt{Q}$.
\end{thm}


\smallskip

\noindent
\underline{\bf Carbon copying onto dirty paper.}
%
A model related to the DPC-FFD is the ``carbon copying onto dirty paper'' of \cite{LapidothCarbonCopying}: in this channel model there are $M$ possible state sequences $S_j$ that can possibly affect in the channel output.
The transmitter has knowledge of each sequence but does not know which one will appear.
Correct decoding must be granted regardless of the state realization and for each of the possible channel output.
\ea{
Y_j^N = X^N + c S_j^N + Z_j^N, \quad j \in [1 \ldots M],
\label{eq:Carbon copying}
}
where $S_j^N$ is an i.i.d. Gaussian sequence for each $j \in [1 \ldots M]$.
In \cite{LapidothCarbonCopying} inner and outer bound to the capacity region are derived but capacity has yet to been determined.

\section{The dirty paper channel with fast fading dirt}
\label{sec:On the capacity of the dirty paper channel with fading dirt and no receiver side information}

We begin by investigating the capacity of DPC-FFD in Fig \ref{fig:WritingOnFadingDirt}: since no closed-form expression for the optimization
in \eqref{eq: capacityWritingOnFadingDirt QS} is available, we derive a novel outer bound that  is expressed solely as a function of the channel parameters.
This outer bound can be approached, for some models, by a simple achievable strategy in which the transmitter to performs Costa pre-coding against the term $c \mu_A S$,
 the average realization of the fading times state.

For the DPC-FFD the term $c \mu_S A$ acts as additional noise, since it is unknown at both the transmitter and the receiver: for this reason in the following we assume that
 $\mu_S=0$.
%
%
%

\begin{thm}{\bf Outer bound and partial approximate capacity for DPC-FFD.\\}
\label{th:No RCSI, general}
Consider the DPC-FFD  in Fig. \ref{fig:WritingOnFadingDirt} and let $h(A)= \f12 \log(2 \pi e \al )$
for some $\al \in [0,1]$, then the capacity $\Ccal$ is upper bounded as
\ea{
\Ccal \leq R^{\rm OUT} = \f12 \log \lb \f {P+1} {c^2 \al} +\f 1 \al \rb +\f 12,
\label{eq:No RCSI OUT, general}
}
and the capacity is to within a gap  $G$ bits/channel-use from  $R^{\rm OUT}$ where
\ea{
G=-\f{\log(\al)}2+\f1 2.
\label{eq:gap DPC-FD}
}
\end{thm}
\begin{IEEEproof}
The proof can be found in App. \ref{app:No RCSI, general}.
\end{IEEEproof}
The gap from capacity in Th. \ref{th:No RCSI, general} can be easily evaluated for some canonical fading distributions.
\begin{lem}{\bf Gap from for some fading distributions.}
\label{lem:Gap from capacity for some canonical fading distributions}

\noindent
$\bullet$  When $A$ is Gaussian distributed with mean $\mu_A$ and unitary variance, the capacity is known to within a gap $G_{\Ncal}$
\ean{
G_{\Ncal}=\f 12.
}

\noindent
$\bullet$ When $A$ is uniformly distributed between $[\mu_A - \f \Delta 2 , \mu_A + \f \Delta 2]$, the capacity can be attained to within a gap $G_{\Ucal}$
\ean{
G_{\Ucal} =- \f12 \log \lb \f{2 \pi e}{12} \rb +\f12 \leq 1.
}

\noindent
$\bullet$ When $A$ is Rayleigh distributed, i.e. $A=\sqrt{U^2+V^2}$ for $U,V \sim \Ncal(0,2/(4-\pi))$ and independent, capacity can be attained to within a gap $G_{\textbf{R}}$ defined as
\ean{
G_{\textbf{R}}=- \f12 \log \lb 1 \rb+\gamma+1+\f12 \leq 2.08,
}
where $\gamma$ is the Euler-Mascheroni constant.

\noindent
$\bullet$ When $A$ is log-normal distributed, i.e. $A=e^{Z_A} e^{-2\mu-\sgs}(e^{\sgs}-1)^{-1}$ for $Z_A \sim (\mu,\sgs)$, capacity can be attained to within a gap $G_{\log}$ defined as
\ean{
G_{\log} = \log\lb e^{\sgs}-1 \rb +\mu+ \sgs+\f12 \leq \mu+2 \sgs -\f12,
}
which is not a finite value for all values of $\mu$ and $\sgs$.
\end{lem}

The result in Th. \ref{th:No RCSI, general} is substantially a negative result since in establishes that, for a number of fading distributions for which $\al$ is close to one, the best strategy is to Costa pre-code against  the mean value of the fading times  the state and treat the term $A_0 S_0$ as additional noise.
This strategy performs very poorly when compared to the full state pre-cancellation and indeed, for any choice of the power $P$, capacity tends to a small constant as the term $c^2$ increases.

Note that the gap $G$ in \eqref{eq:gap DPC-FD} for the log-normal distribution is not bounded: the variance of this distribution  grows exponentially with $\sgs$ while the entropy grows logarithmically with $\sgs$, therefore $\al$  can be made arbitrarily small and $G$ arbitrarily large.
%

In actuality, we expect the outer bound in \eqref{eq:No RCSI OUT, general} to be close to capacity for a larger set of distributions than that for which $\al$ is close to one.
The difficulty in developing a more general result lies in the lack of tighter outer bound.

Note also that this result does not hold for discrete fading distributions and thus does not include extensions of the result in  Th. \ref{thm:outer bound QS DPC-PF optimized 2 values} for the case with no RCSI.

\section{Dirty Paper Channel with Fast Fading Dirt and Receiver Side information}
\label{sec:Dirty Paper Channel with Fading Dirt and Receiver Side information}

We now turn our attention to the DCP-FFD-RCSI: also for this channel  capacity can be obtained from Th. \ref{th:Capacity of the ergodic DPC-PF} but the optimization is extremely hard to express in closed-form.
This  case is significantly harder to study than the case with no receiver fading information because of the distributed way in which transmitter and receiver can cooperate in dealing with
the  term $cAS$.
%
%
%
%
%
%
As an illustrative example, consider the DPC-FFD with no additive noise and in which the state and the input are restricted to take value $\pm 1$, that is
\ea{
Y=X+AS, \quad X,S \in \{-1,1\},
}
while $A$ has any distribution.
Given the cardinality of the input, the capacity of this channel is at most 1 bit/channel-use.
This rate can be attained by setting $X(-1)=X(+1)=1/2$, independent from $S$ and by setting $U=XS$ and independent from $S$ in \eqref{eq: capacityWritingOnFadingDirt QS}.
With this assignment, $U$ can be recovered from the channel output by considering the squared channel output, in fact:
\ea{
(Y^2|A=a)=X^2+a^2 S^2+2 a XS =1+a^2+2 a U,
}
so that $U=(Y^2 -1 -A^2)/2A$, regardless of the distribution of $A$.
This  simple example shows that the maximization in  \eqref{eq: capacityWritingOnFadingDirt QS} might yields some unexpected results.

Given the difficulty of the problem at hand, we are able to make only partial progress in characterizing the capacity of the DPC-FFD-RCSI.
In the following we provide two approximate capacity results for two classes of discrete distributions of $A$:
(i) for the class of discrete distributions in which one of the probability masses is larger or equal to one half
and (ii) for the class of uniform distributions over the discrete set in which points are incrementally spaced apart.
Both results are a generalization of our previous result in Th. \ref{thm:outer bound QS DPC-PF optimized 2 values} and employ a similar inner bound
in which the transmitter simply performs Costa pre-coding against one realization of the fading times the state.
Our contributions is, therefore, to identify a set of channels in which Costa pre-coding is optimal, although it is clear that this coding strategy is not be
capacity achieving in general.

Note that, for the DPC-FFD-RCSI, we again consider the case in which $\mu_S$ is equal to zero: since the receiver has knowledge of $A$, it can subtract $c\mu_S A$ from the channel output.
We also let $\mu_A=0$ for simplicity: the general case is considered in the journal version of this work.

Let's consider first the class of distribution in which there exists an outcome $A=a'$ with $P_A(a')\geq 1/2$: this class of distributions generalizes the distribution
considered in our result in Th. \ref{thm:outer bound QS DPC-PF optimized 2 values}.
For this fading model the transmitter can Costa pre-code against the realization $c a' S$  and obtain full state cancellation for approximatively a portion $P_A(a')$ of the time.
The performance of this strategy can be improved upon letting the channel input be composed of two codewords: one treating the state times fading as noise and one that Costa pre-codes
against $c a' S$. By optimizing over the power allocated to each codeword, one obtains a larger inner bound.
%
%
%
%

\begin{thm}{\bf Approximate capacity for a discrete distribution with a mass larger than half.\\}
\label{thm:approximate capacity discrete mass>1/2}
Consider a DPC-FFD-RCSI in Fig. \ref{fig:WritingOnFadingDirtRCSI} and let $A$ have a discrete distribution $P_A(a')$ with support $\Acal$ where there exists $A=a'$ such that  $P_A(a')\geq 1/2$.
Define moreover
\ean{
& P_A'=P_A(a'), \quad \Po_A'=1-P_A(a') \\
& G = \Po_A' \Ebb[\log(a-a')^2|a \neq a']  \\
& G' = \Po_A' \Ebb \lsb \log \lb \f {(a-a')^2} {a^2} +1 \rb |a \neq a' \rsb,
}
then the capacity $\Ccal$ is upper bounded as
\ean{
\Ccal \leq R^{\rm OUT} = \lcb \p{
\f 12 \log(1+P)+1 & \Po_A' \leq P_A' c^2\\
\f {P_A'} 2 \log(1+P) & P_A' c^2 \leq \Po_A'(P+1) \\
\ \ +\f {\Po_A'}2 \log \lb P {c^2} \rb+1-G/2 &  \\
 \f {P_A'} 2 \log(1+P) +\f 3 2-G/2  & P_A' c^2 > \Po_A'(P+1)
} \rnone
\label{eq:approximate capacity discrete}
}
an the capacity lies to within $G'-G+3$ bits per channel use from $R^{\rm OUT}$.
\end{thm}
\begin{IEEEproof}
The proof can be found in App. \ref{app:approximate capacity discrete}.
\end{IEEEproof}

The result of Th. \ref{thm:approximate capacity discrete mass>1/2} can be evaluated for some discrete fading distributions.
%
\begin{lem}{\bf Gap from for some discrete distributions.}
\label{lem:Gap from capacity for some canonical fading distributions}

\noindent
$\bullet$
When $A$ is distributed according to a geometric distribution, i.e.
\ea{
P_A(k_a+n \Delta)=(1-p)^{n} p, \quad  n \in \Nbb,
}
for some $p \in [0,1],\Delta > 0$ and $p^2 \Delta^2=\po$ (to obtain a unitary variance) ani $k_a=-\Delta(1-p)/p$ (to obtain zero mean),
 %
%
   Th. \ref{thm:approximate capacity discrete mass>1/2}  can be applied for $p \leq 1/2$.
 For this choice of $p$, $A=k_a$ has probability larger than a half and the best strategy  for the transmitter is to Costa pre-code against the sequence $c k_a S$ or otherwise treat the fading times state as noise.
  The value of the outer bound in \eqref{eq:approximate capacity discrete}  depends on the value $G$, while the gap from capacity on $G'$ which are obtained as
  \ea{
  G  & = 2 \sum_{n=1}^{\infty}  \log \lb  n \Delta  \rb p(1-p)^n \geq - (1-p)\log{\Delta^2} \nonumber \\
  G' & =  \sum_{n=1}^{\infty}  \log\lb \f{n^2 \Delta^2 }{ (k_a + \Delta n)^2}+1 \rb p(1-p)^n \leq \f 1 {2 k_a^2} (1-p),
  \label{eq:bounding discrete}
  }
 for which
  \ea{
  G'-G \leq (1-p)(k_a^{-2}+\log \Delta^2 ).
  \label{eq:bound gap}
  }
The gap between inner and outer bound goes to infinite as  $\Delta$ goes to zero: in this regime the channel reduces to the classic DPC with no fading for which the bounding techniques in Th. \ref{thm:approximate capacity discrete mass>1/2} are no longer tight. Note that \eqref{eq:bound gap} goes to infinity as $k_a$ goes to zero, but this is only a consequence of the bounding in \eqref{eq:bounding discrete}.
%

  \noindent
$\bullet$ {\bf Binomial Distribution.}
Consider now the case in which $A$ has a binomial distribution of the form
\ean{
p_A(k_a+n \Delta,N)= \fact{2N}{n}(1-p)^{n} p^{2N-n} , \quad \ n \in [-N \ldots +N],
}
and $2Np(1-p)=\Delta^2$ to maintain the variance unitary and $k_a=-N \Delta p$ to have zero mean.
By simple enumeration we see that for $N>1$ no assignment of $p$ gives a probability mass larger than a half.
For $N=1$ we have only one $p$ which  makes the theorem applicable: $p=1/2$ which corresponds to the probability vector $[1/4 / 1/2 / 1/4]$.
This result extends the case where the probability vector is $[1/2 / 1/2]$ which corresponds to the case it Th. \ref{th:Capacity of the ergodic DPC-PF}.
%
%
%
%
%
%
%
\end{lem}

Another possible extension of the result in Th. \ref{thm:outer bound QS DPC-PF optimized 2 values} is the case in which $A$ is uniformly distributed over a set with more than two elements.
%
%
In the following we indeed show such a generalization: the caveat is that the points in the support of the distribution must be increasingly spaced apart points.
This result is similar in spirit to our result in \cite{rini2014strongFading} for the DPC with slow fading, that is, for the channel in  which a fading coefficient
is randomly drawn from a set of possible values before transmission and is kept constant through the channel transmission.
%
%
The intuitive interpretation of this result is as follows: when two fading value are sufficiently spaced apart, the transmitter cannot exploit the correlation between the two different channel outputs corresponding to the two different fading realizations.
%
%
For this reason the best choice for the transmitter is to Costa pre-code against one realization of the fading times state.
%
%

\begin{thm}{\bf Approximate capacity in the ``strong fading'' regime.\\}
\label{thm:approximate capacity strong}
Consider the case in which $A$ is uniformly distributed over the set
\ea{
\Acal(M) = \lcb a_0 , a_1 \ldots a_M, \quad a_i \in \Rbb \rcb,
}
with $\var(A)=1$ and let $\Delta_{i}$ be the distance between two consecutive  points in $\Acal$,
that is 
\ea{
\Delta_{i+1}=a_{i+1}-a_i, \quad i \in[0 \ldots M-1]
}
and  $\Delta_1>\al$, then, if
\ea{
\Delta_{i+1}^2 \geq (\al  c^2 -1) \sum_{j=1}^{i-1} \Delta_{i+1}^2+2,  \quad i >2
\label{eq:strong fading condition}
}
for some $\al \geq 0$  then an outer bound to capacity is
\ean{
R^{\rm OUT} = \lcb \p{
\f 12 \log \lb 1+\f {P}{c^2+1} \rb+1 & \f {M-1} M \leq \f {c^2} M\\
 \f 1 {2M} \log(1+P)+ & \f {c^2} M \leq \f {M-1}{M}(P+1) \\
 \quad +\f {M-1}{2 M } \log \lb {c^2} \rb+1+\log \al/2 &  \\
 \f 1 {2M} \log(1+P) +1 + \log \al/2  & \f { c^2} M > \f {M-1}{M}(P+1)
} \rnone
}
and the exact capacity lies to within a gap of $\max\{\log(\al)/2-\Gt+3,1\}$ where
\ea{
\Gt = (M-1) \Ebb \lsb \log \lb \f {(a-a')^2} {a^2} +1 \rb |a \neq a' \rsb,
}

\end{thm}
\begin{IEEEproof}
The proof is provided in App. \ref{app:approximate capacity strong}.
\end{IEEEproof}

As an example of Th. \ref{thm:approximate capacity strong} consider the case in which $\al=c^2/(c^2+1)$: in this case the condition in
\eqref{eq:strong fading condition} translates to the set $\Acal(M)$ defined as
\ea{
\Acal(M)= \lcb 0, \Delta_1, c  \Delta_1, c^2 \Delta_1 \ldots c^{M-2} \Delta_1 \rcb - \Delta_0\f{1-c} c,
}
where $\Delta_0$ is determined so that the variance is equal to one, that is
\ea{
\f{\Delta_1^2}{M} \f{1-c^{2M-2} }{1-c^2} - \lb \f{\Delta_1}{M} \f{1-c^{M-1} }{1-c} \rb ^2 =1,
}
which follows from the properties of the geometric series.

Note that Th. \ref{thm:approximate capacity strong} implies that, when $c^2$ is much larger than $P$, then the capacity of the DPC-FFD-RCSI as $1/M$ times the capacity of the
channel without state.

We conclude by providing an outer bound for the case of a continuous fading distribution.
Unfortunately this bound is not tight in general: this reflect the fact that the outer bounding techniques employed so far are too crude to
address this general case.

\begin{thm}{\bf Outer Bound for continuous fading distributions.\\}
\label{thm:approximate capacity continuous}
Consider the case in which $A$ has a continuous distribution with such that there exists a
an interval  $I=[a,b] \subset \Rbb$ with  $P_A(I)\geq 1/2$, let moreover
\ea{
& a' \in [a,b] \ \ST \ P(a')(b-a)=P(I) \nonumber \\
& \Gt  =\int_{\Rbb \setminus I} \log \lb (a-a')^2 \rb \diff P_a,
\label{eq:condition I set}
}
then the capacity $\Ccal$ is upper bounded as
\ean{
\Ccal \leq R^{\rm OUT} = \lcb \p{
\f 12 \log(1+P)+1 & P_A(\Io) \leq P_A(I) c^2\\
 \f {P_A'} 2 \log(1+P)+ & P_A(I) c^2 \leq P_A(\Io)(P+1) \\
 \quad \f {\Po_A'}2 \log \lb P {c^2} \rb+1-\Gt/2 & \\
 \f {P_A'} 2 \log(1+P) +1-\Gt/2  & P_A(I) c^2 > P_A(\Io)(P+1)
} \rnone
}
%
\end{thm}

It is straightforward to verify that the above bound cannot be attained by simply performing Costa pre-coding against  a value of $c a' S$ for some $a'$ of choice:
%
in fact this strategy achieves
\ean{
R^{\rm IN}= \f 12 \log(1+P)- \f 12 \Ebb_A \lsb \log \lb \f{P c^2 }{P+c^2a^2+1} (a-a')^2+1 \rb \rsb \\
\approx \f 12 \log(1+P)- \f 12 \Ebb_A \lsb \log \lb \min \{a^2 c^2,P \} \f{ (a-a')^2}{a^2}+1 \rb \rsb,
}
which goes to zero as $P$ or $c^2$ grows, unless $A$ is mostly concentrated around  $a'\pm 1/c^2$.

\section{Conclusion}
\label{sec:Conclusion}

In this paper we studied a variation of the classic dirty paper channel in which the channel state is multiplied by a fast fading process which is unknown at the transmitter.
We consider both the case in which the decoder has knowledge of the fading  and the case in which it does not.
For this model we derive inner and outer bounds to capacity and bound the difference between the two when possible.
%
%
When fading knowledge in not available at the receiver, the gap between inner and outer bounds is small for a number of classic fading distributions but it is not bounded for others.
%
When fading knowledge is available at the receiver we can characterize capacity for some specific discrete distributions of the fading.
%

\bibliographystyle{IEEEtran}
\bibliography{steBib1}

\onecolumn
\appendix

\subsection{Proof of Th. \ref{th:No RCSI, general}}
\label{app:No RCSI, general}

\noindent
$\bullet$ {\bf Capacity outer bound}
\medskip

Consider the following series of inequalities developed from Fano's inequality
\eas{
&N(R-\ep_N) \\
& \leq  I(Y^N;W)  \\
& \leq  I(Y^N;W|S^N)  \\
&  = h(Y^N|S^N) -h(Y^N|W,S^N, X^N) \\
&  = N \max_j h(Y_j|S_j) -h(Y^N|W,S^N, X^N) \\
&  =  N \max_j \Ebb_{S_j} \lsb h(X_j+c s A_j +Z_j)\rsb -h(Y^N|W,S^N, X^N) \\
& \leq \f N2 \Ebb_{S}  \lsb \log 2 \pi e \lb P+c^2 s^2 + 1 \rb \rsb -h(Y^N|W,S^N, X^N)
\label{eq:bound +H 1} \\
& \leq \f N2  \log  2 \pi e \lb P+c^2 + 1 \rb -h(Y^N|W,S^N, X^N),
\label{eq:bound +H 2}
}{\label{eq:bound +H}}
where \eqref{eq:bound +H 1} follows from the GME property given that $A \perp X$ and $\var[A]=1$ by definition while \label{eq:bound +H 2} follows from Jensen's inequality
and from the fact that $\Ebb[S]=0$.
Note that the mean of $A$ does not influence this bound.

For the term $-h(Y^N|W,S^N, X^N)$ we have:
\eas{
 &-h(Y^N|W,S^N, X^N) \\
 & = -h(c S^N A^N+Z^N|W,S^N, X^N) \\
& = -h(c S^N A^N+Z^N|S^N)
\label{eq:bound -H 1} \\
& = -N h(c S_j A_j+Z_j|S_j)
\label{eq:bound -H 2}  \\
& \leq -N H \lb S_j A_j|S_j \rb -N \log |c|,
}{\label{eq:bound -H}}
where \eqref{eq:bound -H 1}  follows from the Markov Chain $c S^N A^N+Z^N-S^N-W,X^N$, \eqref{eq:bound -H 1} from the fact that $A_i,S_i$ and $Z_i$ are iid RV.

The term $-h(S_j,A_j|S_j)$ can be rewritten as
\ean{
& -h \lb S_j A_j|S_j \rb
= - h(A_j) -　\Ebb_S \lsb \f12 \log(s^2) \rsb = - h(A_j) +　\f  \gamma 2 ,
}
where $\gamma$ is the Euler's constant $\gamma \approx 0.577$.
Note that the derivation holds for $A$ both continuous or discrete.

%
Combining the bounds in \eqref{eq:bound +H} and \eqref{eq:bound -H} we obtain the expression in \eqref{eq:No RCSI OUT, general}.

\bigskip

\noindent
$\bullet$ {\bf Capacity inner bound}
\medskip

For the inner bound, we consider Costa's dirty paper coding strategy to pre-cancel $\mu_A S$ while disregarding the remaining randomness in the fading.
This strategy attains
\ean{
R^{\rm IN}
& =I(Y;U|A)-I(U;S)\\
& =H(U|S)-H(U|Y).
}
Considering now the assignment in which $X$ and $U$
\ean{
X & \sim  \Ncal(0,P), \\
U & =X+k S,
}
which attains
\ea{
R^{\rm IN} \geq \f 1 2 \log \lb   \f {P} {P+k^2-\f{(P+k c \mu_A)^2}{P+c^2(1+\mu_A^2)+1}} \rb
}
by upper bounding $h(U|Y)$ using the GME property.
The optimal choice of $k$ is
\ea{
k^*=\f {P } {P+1+c^2} c \mu_A
}
%
%
which achieves
\ea{
R^{\rm IN} \geq \f 12 \log \lb 1 +\f P {c^2+1} \rb,
\label{eq:NCSI S Gaussian IN}
}
as expected.
%

\bigskip
\noindent
$\bullet$ {\bf Gap between inner and outer bound}
\medskip

By comparing the outer bound expression in \eqref{eq:No RCSI OUT, general}and the inner bound expression in \eqref{eq:NCSI S Gaussian IN} have that the difference in the two expressions
is
\eas{
G
& = R^{\rm OUT} - R^{\rm IN}  \\
& =\f 12 \log 2 \pi e (P+1+c^2) - \f 12 \log (2 \pi e c^2 \al ) +　\f  \gamma 2 \nonumber  \\
& \quad - \lb \f 12 \log 2 \pi e (P+1+c^2) - \f 12 \log 2 \pi e (c^2 +1) \rb \\
& = \f 1 2 \log \lb  \f {c^2+1}{\al c^2}\rb +　\f  \gamma 2  \\
& \leq \f 1 2 \log \lb  \f {4}{ 3 \al c^2}\rb +　\f  \gamma 2
\label{eq:gap simple 1}\\
& \leq \f 1 2 \log \lb  \f 1 {\al }\rb +　\f  1 2,
}{\label{eq:gap simple}}
where \eqref{eq:gap simple 1} follows from the fact that capacity is known to within $1$ bit for  $c\leq 3$.

Equation \eqref{eq:gap simple} concludes the proof.

\subsection{Proof of Th. \ref{thm:approximate capacity discrete mass>1/2}}
\label{app:approximate capacity discrete}

%
%

\bigskip
\noindent
$\bullet$ {\bf Capacity outer bound}
\medskip
\noindent

Using Fano's inequality we write
\eas{
 N(R- \ep)  & \leq I(Y^N;W|A^N) \\
 & \leq \f N2 \Ebb_A \lsb \log2 \pi e(P+a^2 c^2 +2 |c||a|\sqrt{P}  +1 )\rsb -H(Y^N|W,A^N) +\f 12 \\
& \leq \f N2 \Ebb_A \lsb \log2 \pi e(P+a^2 c^2  +1 )\rsb -H(Y^N|W,A^N) +\f 12 \\
& \leq \f N2 \Ebb_A \lsb \log2 \pi e(P+c^2+1 )\rsb-\sum_{a^N \in \Acal^N} H(Y^N|W,A^N=a^N) +\f 12 ,
\label{eq:fano 1}
}{\label{eq:fano}}
where \eqref{eq:fano 1} follows from Jensen's inequality.
Next we derive a bound on $H(Y^N|W,A^N)$ based on the letter-typicality of the sequence $a^n$, defined as
\ea{
\labs \f 1 n N(k|a^N)-P_A(k) \rabs \leq \ep P_A(k)   \quad \forall \ k \ \in \Acal,
\label{eq:typicality cond}
}
where $N(k|a^N)$ is the number of symbols is the sequence $a^N$ which are equal to $k$, i.e.
\ea{
N(k|a^N) = \sum_{j=1}^{N} 1_{\{k=a_j\}}.
}
Accordingly, the $\ep$-typical set $\Tcal_{\ep}^N(P_A)$ is defined as the set of $a^N$ which satisfy \eqref{eq:typicality cond}:
\ea{
\Tcal_{\ep}^N(P_A) = \lcb   a^N, \ \labs \f 1 n N(k|a^N)-P_A(k) \rabs \leq \ep P_A(k) ,  \quad \forall \ k \ \in \Acal \rcb.
}

Using the letter-typicality in \eqref{eq:typicality cond}, we write:
\eas{
& -\sum_{a^N \in \Acal^N } P(a^N) H(Y^N|W,A^N=a^N)  \\
& \leq  -\sum_{a^N \in \Tcal_\ep^N(P_A)} P(a^N) H(Y^N|W,A^N=a^N).
}
%
Let now $\ep \leq  \f {P_{A'}-\f  12} {P_A'}$ so that $N(a'|x^N)>1/2$.
With this provision, we can define the sequence $a'^N$ as a permutation of the sequence $a^N$ where
\begin{itemize}
  \item if $\ao_i\neq a'$, then $a_i=a'$,
  %
  \item if $a_i\neq a'$, then $\ao_i=a'$.
\end{itemize}

\begin{figure}
\centering
\includegraphics[width=.54 \textwidth]{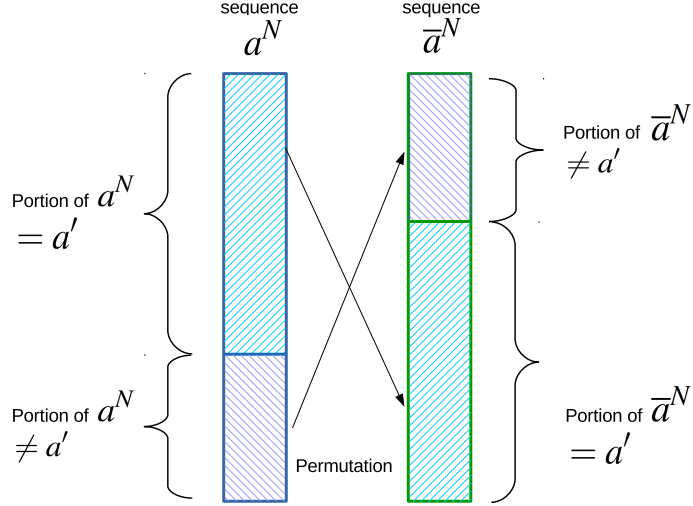}
\caption{
The permutation that generates $a'^N$ from $a^N$ in the proof of Th. \ref{thm:approximate capacity discrete mass>1/2} in App. \ref{app:approximate capacity discrete}.
}
\label{fig:permutation}
\end{figure}

This permutation is also depicted in Fig.  \ref{fig:permutation}: the sequence $a'^N$ is obtained by permuting  the positions $i$ for which $a_i \neq a'$ with some of the positions $j$ for which $a_j = a'$: since $N(a'|x^N)>1/2$, this can always be done.
Note that $N-2 (N- N(a'|a^n))=2 N(a'|a^n) - N$ positions are such that $a_i=\ao_i=a'$.

With this definition of $a'^N$ we next define the equivalent channel output
\ea{
\Yo=X^N+c a'^N S^N + \Zo^N,
\label{eq:equivalent Y >1/2}
}
where $\Zo^N $ has the same marginal distribution of $Z^N$ and any chosen joint distribution with this term.
%

With these definitions in place, we write:
\eas{
&-\sum_{a^N \in \Tcal_\ep^N(P_A)} P(a^N) H(Y^N|W,A^N=a^N)\\
& = - \f12 \sum_{a^N \in \Tcal_\ep^N(P_A)} P(a^N) \lb H(Y^N|W,A^N=a^N)+H(\Yo^N|W,A^N=a'^N) \rb  \\
%
%
& \leq - \f12 \sum_{a^N \in \Tcal_\ep^N(P_A)} P(a^N) \lb H(X^N+c a^N S^N + Z^N,X^N+c a'^N S^N + \Zo^N |W) \rb  \\
& = - \f12 \sum_{a^N \in \Tcal_\ep^N(P_A)} P(a^N)  H \lb c (a^N-a'^N)S^N + Z^N- \Zo^N,X^N+c a'^N S^N + \Zo^N |W \rb   \\
& = - \f12 \sum_{a^N \in \Tcal_\ep^N(P_A)} P(a^N) \lb  H \lb c (a^N-a'^N)S^N + Z^N- \Zo^N \rb + H(\Yo|Y-\Yo,W,S^N,X^N   ) \rb
\label{eq: trick larger 1/2 1} \\
& \leq - \f12 \sum_{a^N \in \Tcal_\ep^N(P_A)} P(a^N) \lb  H \lb c (a^N-a'^N)S^N + Z^N- \Zo^N \rb + H(\Zo^N) \rb  \\
& =- \f12 \sum_{a^N \in \Tcal_\ep^N(P_A)} P(a^N) \lb  H \lb c (a^N-a'^N)S^N + Z^N- \Zo^N \rb +  \f  N  2 \log (2 \pi e )\rb
%
}{\label{eq: trick larger 1/2}}
where \eqref{eq: trick larger 1/2 1}  follows from the fact that $S^N$ and the additive noises are independent from $W$.

Let us now focus solely on the term $1/2 \sum_{a^N \in \Tcal_\ep^N(P_A)} P(a^N) H \lb c (a^N-a'^N)S^N + Z^N- \Zo^N \rb$:
we can make use of the following properties of the typical sets:
\eas{
& P(a^N) \leq \f 1 {2^{n(1+\ep) H(A)}} , \quad a^N \in \Tcal_\ep^N \\
& \labs \Tcal_\ep^N(P_A) \rabs \leq (1-\delta_{\ep})2^{n(1-\ep) H(A)} \\
& N(k|a^N) \leq   N P_A(k)(a)(1-\ep),
}{\label{eq:typical properties} }
for
\ea{
\delta_{\ep} = 2|\Acal|e^{ - n 2 \min_k P_A(k)}.
}
Using the properties in \eqref{eq:typical properties} we now write:
\eas{
& - \f1 2  \sum_{a^N \in \Tcal_\ep^N(P_A)} P(a^N) H \lb c (a^N-a'^N)S^N + Z^N- \Zo^N \rb \\
& \leq  - \f 12 \f 1 {2^{-n(1+\ep) H(A)}}  \sum_{a^N \in \Tcal_\ep^N(P_A)}  H \lb c (a^N-a'^N)S^N + Z^N- \Zo^N \rb   \\
& \leq - \f 12 \f 1 {2^{-n(1+\ep) H(A)}}  \sum_{a^N \in \Tcal_\ep^N(P_A)}  \sum_{i=1}^N \lb  H \lb c (a_i-\ao_i)S_i + Z_i- \Zo_i  \rb \rb.
{\label{eq: trick larger 1/2V1 1}}
}{\label{eq: trick larger 1/2V1}}
We would now wish to change the summation in the right hand side of  \eqref{eq: trick larger 1/2V1 1} from $i \in [1 \ldots N]$ to $k \in \Acal$.
To do so we need to remember how $a'^N$ was defined: $a_i-\ao_i$ can take values: $a'-a$, $a-a'$ and $0$.
Since the entropy term $H \lb c (a_i-\ao_i)S_i + Z_i- \Zo_i  \rb $ is not affected  by the sign of $|a_i-\ao_i|$, we conclude that there are $2(N-N(a'|a^N))$ times in which we have
$H \lb c (a'-k)S_i + Z_i- \Zo_i  \rb $ for some $k \neq a'$ and $2 N(a'|a^N) - N$ terms with value $H \lb Z_i- \Zo_i  \rb$.
Additionally, for a given $k$,  $H \lb c (a'-k)S_i + Z_i- \Zo_i  \rb $ appears $N(k|a^N)$ times.

With these observations we now write
\eas{
& - \f 12 \f 1 {2^{-n(1+\ep) H(A)}}  \sum_{a^N \in \Tcal_\ep^N(P_A)}  \sum_{i=1}^N \lb  H \lb c (a_i-\ao_i)S_i + Z_i- \Zo_i  \rb \rb  \\
& = - \f 12 \f 1 {2^{-n(1+\ep) H(A)}}  \sum_{a^N \in \Tcal_\ep^N(P_A)}  \sum_{k \in \Acal \setminus a'} 2  N(k|a^N) H \lb c (a'-k)S_i + Z_i- \Zo_i  \rb \nonumber\\
& \quad \quad - \f 12  (2 N(a'|a^N) - N) H \lb Z_i- \Zo_i  \rb.
}{\label{eq:typicality 1}}
We can now choose the joint distribution between $Z_i$ ani $\Zo_i$ to simplify the bound above: for simplicity we choose $Z^N=\Zo^N$.
%
%
With this choice, we can write
\eas{
& - \f 12 \f 1 {2^{-n(1+\ep) H(A)}}  \sum_{a^N \in \Tcal_\ep^N(P_A)}  \sum_{k \in \Acal \setminus a'} 2  N(k|a^N) H \lb c (a'-k)S_i + Z_i- \Zo_i  \rb  \nonumber\\
& \quad \quad   - \f 12 (2 N(a'|a^N) - N) H \lb Z_i- \Zo_i  \rb\\
& =  - \f 12 \f 1 {2^{-n(1+\ep) H(A)}}  \sum_{a^N \in \Tcal_\ep^N(P_A)}  \sum_{k \in \Acal \setminus a'} 2  N(k|a^N) H \lb c (a'-k)S_i\rb   - \f N 4 \log ( 4 \pi e)  \\
& =  - \f 12 \f 1 {2^{-n(1+\ep) H(A)}}  \sum_{a^N \in \Tcal_\ep^N(P_A)}  \sum_{k \in \Acal \setminus a'} 2  N(k|a^N) \f 12 \log (2 \pi e c^2 (a'-k)^2 )   - \f N 4  \log (4  \pi e)  \\
& =  - \f 1 {2^{-n(1+\ep) H(A)}}  (1-\delta_{\ep})2^{n(1-\ep) H(A)} \sum_{k \in \Acal \setminus a'}  N(k|a^N) \f 12 \log ( 2 \pi e c^2 (a'-k)^2 )   \nonumber \\
& \quad  \quad - \f N 4  \log ( 4 \pi e)  \\
& =  - \f 1 {2^{-n(1+\ep) H(A)}}  (1-\delta_{\ep})2^{n(1-\ep) H(A)}(1-\ep) N  \sum_{k \in \Acal \setminus a'} P_A(k)(a)\f 12 \log (2 \pi e c^2 (a'-k)^2 ) \nonumber \\
& \quad  \quad  - \f N 4  \log (4  \pi e).
}{\label{eq:typicality 2}}
When  $N$ is sufficiently large and $\ep$ sufficiently small, we then have that
\eas{
& - H(Y^N|W,A^N) \\
& \quad  \leq - \sum_{k \in \Acal \setminus a'} P_A(k)(a)\f 12 \log ( 2 \pi e c^2 (a'-k)^2 )   - \f N 4  \log ( 4 \pi e)  - \ep_{\rm all}
\label{eq:epsilon all 1}\\
& \quad  \leq - \f {N \Po_{A}'} 2 \log c^2 - \f {N G}  2  - \f N 2  \log (4 \pi e)- \ep_{\rm all} ,
}{\label{eq:epsilon all}}
for some $\ep_{\rm all}$ that goes to zero as $N \goes \infty$.

Using the bound in \eqref{eq:epsilon all} in \eqref{eq:fano 1} and for some $\ep_{\rm all}$ sufficiently small, we obtain
\eas{
R^{\rm OUT}
& =  \f12 \log \lb 2 \pi e(P+c^2+1 )\rb  - \f{\Po_A'} 2 \log  c^2  -\f G2- \f 1 4 \log(2 \pi e) + \f 12 \\
&  \leq \f12 \log \lb P+c^2 +1 \rb  - \f{\Po_A'} 2 \log (c^2 )-\f G2 + 1,
} {\label{eq:outer before optimization}}
We next optimize the above expression over the parameter $c^2$ over the set $[0,c^2]$ since capacity must be decreasing in $c$.
The optimal value of $c^2$ in \eqref{eq:outer before optimization} is
\ea{
\lb c^2 \rb^*=\min \lcb \f {\Po_A'}{P_A' }(1+P),c^2 \rcb.
}
When $P_A'  c^2 \geq \Po_A' (1+P)$ this optimization yield the tighter outer bound than the original outer bound in \eqref{eq:outer before optimization}
\ea{
& \lnone R^{\rm OUT} \rabs_{P_A'  c^2 \geq \Po_A' (1+P)} \\
& \quad \quad   = \f {P_A'} 2   \log(1+P)+ \f 12 h_2(P_A') - \f  G 2 + 1 \\
& \quad \quad \leq \f {P_A'} 2   \log(1+P) - \f  G 2 + \f 3 2,
}
where $h_2(x)$ indicates the binary entropy.
%
%
so that the overall outer bound can be further simplified as
\ea{
R^{\rm OUT}=\lcb\p{
 \f12 \log \lb P+c^2 +1 \rb  & \\
  \quad - \f{\Po_A'} 2 \log (c^2 )-\f G2 + 1 & P_A' c^2 \leq \Po_A'(P+1) \\
 \f {P_A'} 2 \log(1+P) - \f G  2  +\f 3 2  & P_A' c^2 > \Po_A'(P+1).
}\rnone
\label{eq:outer bound >1/2}
}
%

\noindent
\bigskip
$\bullet$ {\bf Capacity inner bound}
\medskip
\noindent

For the inner bound consider the simple scenario in which the transmitter Costa pre-codes against the realization $c a'S$, which occurs more than half of the time.
That is, consider the assignment
\ean{
& X \sim \Ncal(0,P) \\
& U = X+ \f P {P+1} a' c S, \quad U \perp X.
}
The attainable rate of this scheme is
\eas{
R^{\rm IN}
& = \Ebb_A \lsb \lsb I(Y;U|A)-I(U;S)\rsb^+ \rsb \\
& \geq \f {P_A'} 2  \log(1+P) + \sum_{\Acal, a\neq a'}  \f {P_A(a)}2 \log \lb \f {(1+c^2 a^2+P)(1+P)}{P_A' c^2 (a-a')^2+P+c^2a^2+1} \rb,
}{\label{eq:inner bound al }}
the latter term is bounded as
\eas{
& \sum_{\Acal, a\neq a'}  \f {P_A(a)} 2 \log \lb \f {(1+c^2 a^2+P)(1+P)}{P c^2 (a-a')^2+P+c^2a^2+1} \rb \\
%
&  = \sum_{\Acal, a\neq a'}  \f {P_A(a)} 2 \log \lb 1+P \rb - \f {P_A(a)} 2 \log \lb \f{P a^2 c^2 }{P+c^2a^2+1} \f{(a-a')^2}{a^2}+1 \rb \\
&  \geq \sum_{\Acal, a\neq a'}  \f {P_A(a)} 2 \log \lb P \rb - \f {P_A(a)} 2 \log \lb \f {\min \{P, a^2 c^2 \}} 2  \f{(a-a')^2}{a^2}+1 \rb \\
& \geq  \sum_{\Acal, a\neq a'}  - \f {P_A(a)} 2 \log \lb \min \lcb 1,\f { a^2 c^2}{P} \rcb \f{(a-a')^2}{a^2}+\f 1 P \rb \\
& \geq  \sum_{\Acal, a\neq a'}  - \f {P_A(a)} 2 \log \lb \min \lcb 1,\f { a^2 c^2}{P} \rcb \f{(a-a')^2}{a^2}+1  \rb \\
& \geq  \sum_{\Acal, a\neq a'}  - \f {P_A(a)} 2 \log \lb \f{(a-a')^2}{a^2}+1  \rb = -G'.
\label{eq:inner bound only binning >1/2 1}
}{\label{eq:inner bound only binning >1/2}}

This attainable rate can be improved upon by using two codewords: one that treats the interference  as noise.
We can assign power $\al$ to one codeword and power $\alb=1-\al$ to the other and successively optimize over the power assigned to each codeword.
This yield the achievable rate
\eas{
 R^{\rm IN}
& = \max_{\al \in [0,1]} \Ebb_A  \lsb \f 12 \log \lb 1 + \f {\al P}{1+c^2a^2+\alb P}\rb \rnone \nonumber \\
& \quad \quad \lnone + \f {P_A'} {2} \log \lb 1 +\alb P  \rb   + \sum_{\Acal, a\neq a'}  \f {P_A(a)}2 \log \lb \f {(1+c^2 a^2+\alb P)(1+\alb P)}{P_A' c^2 (a-a')^2+\alb P+c^2a^2+1} \rb \rsb
\\
& \geq \max_{\al \in [0,1]} \Ebb_A  \lsb \f 12 \log \lb 1 + \f {\al P}{1+c^2a^2+\alb P}\rb + \f {P_A'} {2} \log \lb 1 +\alb P  \rb \rsb - \f {G'} 2
\label{eq:bound negative entropy term} \\
& \geq \max_{\al \in [0,1]} \f 12 \log \lb 1 + \f {\al P}{1+c^2+\alb P}\rb + \f {P_A'} {2} \log \lb 1 +\alb P  \rb  - \f {G'} 2,
}
where \eqref{eq:bound negative entropy term} follows from the fact that the bound in \eqref{eq:inner bound only binning >1/2} holds for any $P$.

the optimal value of $\alb P $ is then
\ea{
\alb^* P=\max  \lcb \min \lcb \f{P_A'}{\Po_A'} c^2-1 , P\rcb, 0 \rcb,
}
so that, when $\Po_A' \leq P_A' c^2 \leq \Po_A'(P+1)$ we have
\eas{
R^{\rm IN}=  \f 12 \log (P+c^2 +1) - \f {\Po_A'} 2 \log \lb c^2  \rb - h_2(P_A') \\
\geq \f 12 \log (P+c^2 +1) - \f {\Po_A'} 2 \log \lb c^2  \rb -1 - \f {G'}{2}.
}
Finally, we have shown the achievability of the outer bound
\ea{
R^{\rm IN}=\lcb \p{
 \f 12 \log \lb 1 + \f{P} {1+c^2} \rb & \Po_A' \leq P_A' c^2 \\
%
\f 12 \log (P+c^2 +1)   &  \Po_A' \leq P_A' c^2 \leq \Po_A'(P+1) \\
\quad - \f {\Po_A'} 2 \log \lb c^2  \rb -1 - \f {G'}{2} \\
 \f {P_A'} 2 \log(1+P) - 1 - \f {G'}{2}   & P_A' c^2 > \Po_A'(P+1)
}\rnone
\label{eq:inner bound >1/2}
}

\bigskip
\noindent
$\bullet$ {\bf Gap between inner and outer bound}
\medskip

A gap between inner and outer bound of 3 bits in the interval $P_A' c^2 > \Po_A' $ can be obtained by comparing the two expressions in \eqref{eq:inner bound >1/2} and
\eqref{eq:outer bound >1/2} in the cases i) $\Po_A' \leq P_A' c^2$ , ii) $\Po_A' \leq P_A' c^2 \leq \Po_A'(P+1)$  and iii) $P_A' c^2 > \Po_A'(P+1)$.

For the case in which $\Po_A' \leq P_A' c^2$ we have that $c^2 \leq 1$ so that the capacity can be approached to within 1 bit by treating the
interference as noise with a variance partially known at the receiver.

In the other two cases the gap is at most $-\f G 2 +\f {G'} 2 +3$.

\subsection{Proof of Th. \ref{thm:approximate capacity strong}}
\label{app:approximate capacity strong}

\bigskip
\noindent
$\bullet$ {\bf Capacity outer bound}
\medskip
\noindent

%
We proceed in the bounding from Fano's inequality up to \eqref{eq:fano} in App. \ref{app:approximate capacity discrete}.
%

We next wish to construct now a sequence $a'^N$ from $a^N$ as done it the proof of Th. \ref{thm:approximate capacity discrete mass>1/2}: for this proof we actually need to construct  $M-1$ auxiliary sequences, $a_{(k)}^N$, obtained as
\ea{
a_{(k)}^N = \lcb  a_i=\al_j \implies  a_{(k),i}=\al_{ {\rm mod}(k+j,M)}, \ \forall j\in [1 \dots ]\Acal \rcb  \quad k \in [0 \ldots M-1].
}

Accordingly we define $Y_{(k)}^N$ as the channel output obtained when the fading sequence is $a_{(k)}^N$ as in \eqref{eq:equivalent Y >1/2},
\ea{
Y_{(k)}^N= X+a_{(k)}^N S^N + Z_{(k)}^N.
\label{eq:equivalent Yk}
}
Note that, as in the proof of Th. \ref{thm:approximate capacity discrete mass>1/2}, we can associate a different noise to each $Y_{(k)}^N$  in \eqref{eq:equivalent Yk} and later choose the joint distribution among these noise terms.
Since the symbols are equiprobable, we have that $P(Y^N|W, A^N=a_{(0)}^N)=P(Y^N|W, A^N=a_{(k)}^N)$ for all $k$.
Additionally, given the definition of typicality in \eqref{eq:typicality cond}, if $a^N \in \Tcal_{\ep}^N$, we have that also  $a_{(k)}^N \in \Tcal_{\ep}^N$.
As a last definition, let  $Y_{(k)}^N(j)$ be the subset of position of $Y_{(k)}^N(j)$ in which $a_{(k),j}=\al_j$ for $j \in [1 \ldots M]$ in the chosen ordering of $\Acal$, that is
\ea{
Y_{(k)}^N(j) = \lcb  Y_{(k),i}(j), \ \ST,  a_{(k),i}=\al_j , \ i \in [1 \ldots N]\rcb, \quad \forall j \in [1 \ldots M],
\label{eq:definition Yj}
}
Accordingly $Y_{(0)}(m)$ is the subsets of channel outputs in which $a_j=x_m$ and $Y_{(k)}({\rm mod}(m+k))$ are the same subsets of outputs but in which $a_j=x_{{\rm mod}(m+k)}$.

\medskip

A first part of the proof involves extending the bounding in \eqref{eq: trick larger 1/2} to the case of any number of passible fading realization $M=|\Acal|$.
This derivation involves a recursion which  we illustrate this using the case $M=3$:  the general case is inferred from this derivation.
We shall continue the derivation of the outer bound from \eqref{eq:fano 1} and focusing on the bounding of the term $-H(Y^N|W,A^N)$.

\medskip
$\bullet$ {\bf Case for $M=3$}
\smallskip

Consider $M=3$ and $\Acal=\{a_1,a_2,a_3\}$ for some ordering of the elements in $\Acal$ and, as in \eqref{eq:typicality 1} note that
\eas{
-H(Y^N|W,A^N)
& \leq - \sum_{a^N \in \Tcal_\ep^N(P_A)} P(a^N)   H(Y^N|W,a^N) \\
& =  - \sum_{a^N \in \Tcal_\ep^N(P_A)} \f 1 {3^N}   H(Y^N|W,a^N)  +\ep_{\rm all}
}
%

%
\begin{figure}
\centering
\includegraphics[width=.8 \textwidth]{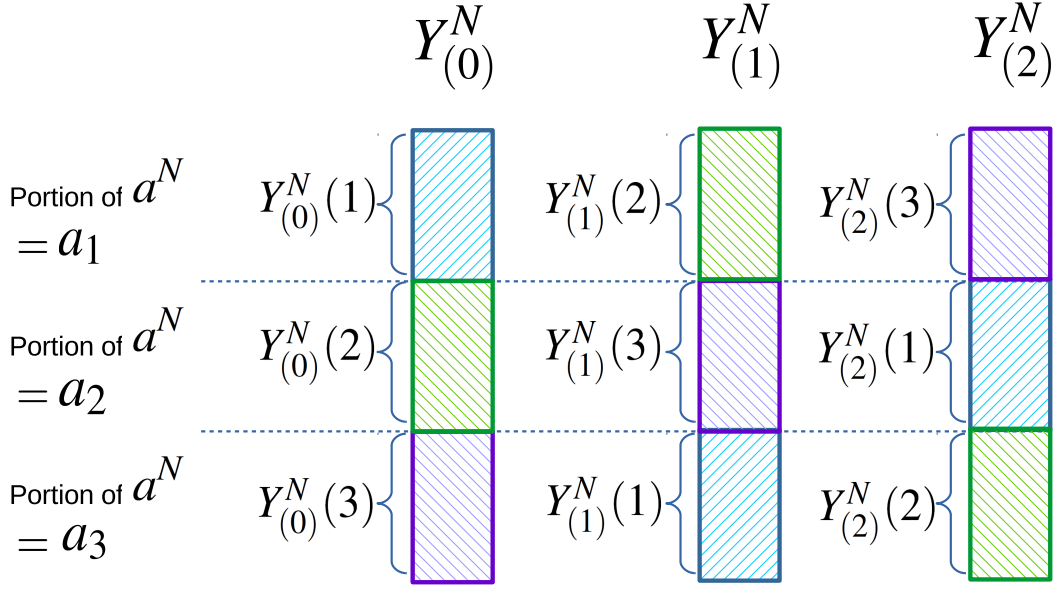}
\vspace{-2.5 cm}
\caption{
An illustration of the The sequences $Y_{(k)}^N$ and the subsequences $Y_{(k)}^N(j)$ for $k,j \in \{1,2,3\}$  in App. \ref{app:approximate capacity strong}
}
\label{fig:permutationUniform}
\end{figure}
The sequences $Y_{(k)}^N$ and the subsequences $Y_{(k)}^N(j)$ for $k,j \in \{1,2,3\}$ are illustrated it Fig. \ref{fig:permutationUniform} from which we see that
 $Y_{(0)}(1)$, $Y_{(1)}(2)$ and $Y_{(2)}(3)$ are obtained from the same set of $X$s, $S$s and $Z$s but different fading value.

For this reason we can write
\eas{
& -H(Y^N|A^N=a^N)  \\
& = - \f 1 3   \lb H(Y^N|W, A^N=a_{(0)}^N) + H(Y^N|W, A^N=a_{(1)}^N) + H(Y^N|W,A^N=a_{(2)}^N)\rb  \\
& \leq - \f 1 3 \lb H(Y_{(0)}^N,Y_{(1)}^N,Y_{(2)}^N|W) \rb  \\
& = - \f 1 3 \lb H(Y_{(0)}^N(1),Y_{(0)}^N(2),Y_{(0)}^N(3),Y_{(1)}^N(1),Y_{(1)}^N(2),Y_{(1)}^N(3),Y_{(2)}^N(1),Y_{(2)}^N(2),Y_{(2)}^N(3) |W) \rb,
\label{eq:example 3 v1}
}{\label{eq:example 3}}
where \eqref{eq:example 3 v1} follows form the fact that the transformation of variables has Jacobian one.

Using the definition of $Y_{(k)}^N(j)$ in \eqref{eq:definition Yj} we conclude that the vector
\ea{
\lsb  Y_{(1)}(2) - Y_{(0)}^N(1), \ Y_{(0)}^N(2)- Y_{(2)}(1) ,  \ Y_{(2)}(2) - Y_{(1)}(1) \rsb,
}
is a permutation of the vector
\ea{
c (a_2-a_1)S^N+ \Zt_{21}^N,
}
where  $\Zt_{21}^N$  is a permutation of  the terms
\ea{
\lsb Z_{(0)}(2)-Z_{(2)}(1),  \ Z_{(1)}(2)-Z_{(0)}(1), \  Z_{(2)}(2)-Z_{(1)}(1) \rsb.
\label{eq:Z21}
}

We then have
\eas{
& - 3 H(Y^N|W,A^N=a^N)   \\
& \leq H(Y_{(0)}^N(2),Y_{(0)}^N(3),Y_{(2)}^N(2),Y_{(2)}^N(3),Y_{(3)}^N(2),Y_{(3)}^N(3)|W,(c (a_2-a_1)S^N)+\Zt_{21}^N|W)  \nonumber \\
& \quad  \quad - H(c (a_2-a_1)S^N+\Zt_{21}^N|W),
%
\label{eq:uniform 3 first group}
}{\label{eq:uniform 3 first group all}}
where  \eqref{eq:uniform 3 first group} follows from the fact that this transformation has unitary Jacobian.
Consider now the vector
\ea{
\lsb Y_{(2)}^N(3)-Y_{(1)}(2), \ Y_{(1)}(3) - Y_{(0)}^N(2),  \ Y_{(0)}(3) - Y_{(2)}(2) \rsb,
}
which is again a permutation of the vector
\ea{
c (a_3-a_2) S^N+ \Zt_{32}^N,
}
where $\Zt_{32}^N$ is a permutation of the noise vector
\ea{
\lsb Z_{(2)}^N(3)-Z_{(1)}(2), \ Z_{(1)}(3) - Z_{(0)}^N(2),  \ Z_{(0)}(3) - Z_{(2)}(2)  \rsb.
}
With this definition we can write
\eas{
 & - 3 H(Y^N|W,A^N=a^N)    \\
& \leq  H( c(a_3-a_2) S^N+\Zt_{32}^N | c (a_2-a_1)S^N +\Zt_{21}^N)- H(c (a_2-a_1)S^N +\Zt_{21}^N) \nonumber \\
&  \quad \quad - H( Y_{(0)}(3),Y_{(1)}(3),Y_{(2)}(3)| c (a_2-a_1)S^N +\Zt_{21}^N, c(a_3-a_2) S^N+\Zt_{32}^N, W) \\
&  \leq  H( c(a_3-a_2) S^N+\Zt_{32}^N | c (a_2-a_1)S^N +\Zt_{21}^N)- H(c (a_2-a_1)S^N +\Zt_{21}^N)- H(\Zt_3^N),
\label{eq: M3 sigle letter}
}
where $\Zt_3^N$ is a permutation of the noise terms
\ea{
\lsb  Z_{(0)}(3), \ Y_{(1)}(3), \ Y_{(2)}(3) \rsb.
}
%
%

The expression in  \eqref{eq: M3 sigle letter} is composed of vectors of independent terms, but the distribution of $\Zt_{21}^N$ and $\Zt_{32}^N$ might not be identical, since we
haven't chosen a joint distribution between the noise terms.
At this point in the proof we can sen the noises to be independent so that
\ea{
\Zt_{21,i}, \  \Zt_{32,i } \sim \Ncal(0,2),
}
and iid for all $i\in [1 \ldots N]$.
We can now evaluate the terms in \eqref{eq: M3 sigle letter} for this assignment as
\ea{
H(c (a_2-a_1)S^N +\Zt_{21}^N)= \f N 2 \log 2 \pi e \lb c^2 \Delta_1^2 +2   \rb ,
}
and
\eas{
& H(c(a_3 -a_2) S^N +\Zt_{32}^N|c (a_2-a_1) S^N +\Zt_{21}^N)\\
& = N H \lb c^2(a_3 -a_2)(a_2-a_1) \lb 1 -\f{c (a_2-a_1) }{ c^2 (a_2-a_1)^2 +1 } S\rb  + \Zt_{32} - \f{c^2 (a_3 -a_2)(a_2-a_1)}{c^2 (a_2-a_1)^2 +1 } \Zt_{21}   \rb,
}
where we have $\Zt_{32}$ and $\Zt_{21}$ are zero mean Gaussian with variance two.
which can be further simplified as
\eas{
 & H(c(a_3 -a_2) S +\Zt_{32}|c (a_2-a_1) S +\Zt_{21}) \\
 &=  H(c\Delta_2 S +\Zt_2|c \Delta_1 S +\Zt_1)  \\
& = \f 1 2 \log  2 \pi e \lb c^2 \Delta_2^2 +2 - \f{c^4 \Delta_1^2 \Delta_2^2}{c^2 \Delta_1^2+2}\rb \\
%
%
%
& = \f 12 \log 2 \pi e \lb \f {2 c^2 (\Delta_2^2+\Delta_1^2)+4}{c^2 \Delta_1^2+2} \rb\\
& = \f 12 \log \lb \f {c^2 (\Delta_2^2+\Delta_1^2)+2}{c^2 \Delta_1^2+2} \rb + \log 2 \pi e
%
}{\label{eq:kpsyw}}
The conditions in \eqref{eq:strong fading condition} for $M=3$ become
%
\eas{
& \Delta_1^2 \geq  \al,\\
& \Delta_2^2 \geq  (\al c^2-1) \Delta_1^2
\quad \implies \quad \Delta_2^2+\Delta_1^2 \geq \al c^2 \Delta_1^2,
\label{eq:cond M=3}
}
for some $\al >0$ so that we can write
\ea{
- 3 H(Y^N|A^N=a^N) \leq 2 \lb  - \f N 2 \log  2 \pi e(c^2) - \f N 2   \log \al  \rb- \f N 2 \log 2 \pi e.
}

Note that when \eqref{eq:cond M=3} holds, then the entropy term $H(Y|a^N,W)$ no longer depends on $a^N$ and thus we have that
\ea{
-\sum_{a^N  \in \Acal^N} P(a^N) H(Y^N|A^N=a^N)  \leq -  \f {M-1}{2 M }  \lb \log 2 \pi e c^2 - \f 1 2 \log   \al  \rb - \ep_{\rm all},
\label{eq:epsilon all 2}
}
for some $\ep_{\rm all}$ which goes to zero as $N$ goes to infinity.
Equation \eqref{eq:epsilon all 2} follows, similarly to \eqref{eq:epsilon all}, from the fact that the typical set $\Tcal_{\ep}^N(P_A)$ contains most of the probability and that the sequences in the typical set have a sample probability close to the $P_A$.

\medskip
$\bullet$ {\bf Case for general $M$}
\smallskip

The derivation for the case $M=3$ can be extended to the general case by generalizing the bounding in \eqref{eq:example 3}, \eqref{eq:uniform 3 first group all} and \eqref{eq:kpsyw} to any $M$.
Typicality, as in \eqref{eq:epsilon all 2}, can be invoked to obtain a bound on the term $H(Y^N|W,A^N)$.
%
The bound in \eqref{eq:example 3}

\begin{itemize}
  \item produce $M-1$ sequence $a_{(k)}^N$ and the corresponding sequences $Y_{(k)}^N$ so that
  \eas{
  -H(Y^N|W,a^N)
  & =-\f 1 M  \sum_{k=1}^N H(Y^N_{(k)}|W) \\
  & \leq -\f 1 M  H(Y^N_{(0)}, \ldots,Y^N_{(M-1)}|W).
  \label{eq:minus entropy M}
  }
  This expands on the bounding in \eqref{eq:example 3}
  \item
  Obtain the term $(a_2-a_1)S^N+\Zt_{21}$ as combination of the terms $Y_{(k)}(2)-Y_{(\mod(k+M-1,M))}(1)$ from the entropy term in \eqref{eq:minus entropy M}: this transformation is composed of a circular matrix and an identity matrix which can be shown to have unitary determinant.
  This term can be removed it from the term in  using the definition of conditional entropy and bounded as $-N/2 \log (2 \pi e c^2) + 1/2 \log(\al)$
  This generalizes the passage in \eqref{eq:uniform 3 first group all}.

  \item  Successively remove the terms $\Delta_i S^N+\Zt_{i(i-1)}$ so that
  \ea{
  -H(Y^N|A^n=a^n) \leq \f 1 M  \sum_{i=1}^{M-1} N H(\Delta_i S + \Zt_i | \Delta_1 S +\Zt_1 \ldots \Delta_{i-1} S +\Zt_{i-1}) - H(\Zt_M),
  \label{eq:Y decomposed case M}
  }
where $\Zt_{i,i+1}$ is defined analogously to $\Zt_{21}$ in \eqref{eq:Z21}.

Each term $H(\Delta_i S + \Zt_i | \Delta_1 S +\Zt_1 \ldots \Delta_{i-1} S +\Zt_{i-1})$ in \eqref{eq:Y decomposed case M} can be evaluated as
\ea{
H(\Delta_i S + \Zt_i | \Delta_1 S +\Zt_1 \ldots \Delta_{i-1} S +\Zt_{i-1}) = \f 12 \log \lb \f{ c^2 (\sum_{j=1}^i \Delta_j^2)+2}{c^2 \lb \sum_{j=1}^{i-1} \Delta_j^2 \rb +2} \rb.
}
This term, under the condition in \eqref{eq:strong fading condition} can be bounded as $1/2 \log 2 \pi e c^2 + 1/2 \log \al$.

This generalizes the bounding in  \eqref{eq:kpsyw}.
\end{itemize}

\bigskip

With the above recursion we come to the outer bound
\ea{
R^{\rm OUT}  & = \f 12 \log (1+(1+\mu_A^2)c^2 +P) - \f {M-1}{2M} \log((1+\mu_A^2) c^2) + \f 1 2 \log (\al)- \f {M-1} {2M} + \f 12
}
This  expression correspond to the expression in  \eqref{eq:outer before optimization} in the proof of Th. \ref{thm:approximate capacity discrete mass>1/2},
consequently in can be optimized over $c$ as such said expression.
This results in the outer bound

\ea{
R^{\rm OUT}=\lcb\p{
 \f12 \log \lb P+c^2(1+\mu_A^2) +1 \rb  - \f {M-1} {2M} \log (c^2  (1+\mu_A^2)  ) & \\
 \quad \quad - \f{M-1} {2M} \log (\al) + \f 1 2 & \f 1 M c^2 (1+\mu_A^2) \leq \f {M-1} M (P+1) \\
 \f 1  {2M} \log(1+P) - \f{M-1} {2M} \log (\al)  +\f 3 2  & \f 1 M c^2 (1+\mu_A^2) > \f {M-1} M (P+1).
}\rnone
\label{eq:outer bound M}
}

%
%

\noindent
\bigskip
$\bullet$ {\bf Capacity inner bound}
\medskip
\noindent

For the inner bound, consider the case in which the transmitter pre-codes  against one of the realizations of the state times the fading.
Let such realization be $a' S^N$ so that we attain the rate
\ea{
R^{\rm IN} \geq  \f 1 {2 M} \log(1+P) -  \f 1 {2 M} \sum_{\Acal, a\neq a'}  \log \lb \f{P c^2 (a-a')^2}{P+c^2 a^2 +1}+1  \rb .
\label{eq: binning only strong}
}
as in \eqref{eq:inner bound only binning >1/2}.
Using the definition of $\Gt$ in
\ea{
R^{\rm IN} \geq   \f 1 {2 M} \log(1+P) -  1 .
}

By combining the scheme in \eqref{eq: binning only strong} with the scheme that treats the fading-times-state as noise we attain the bound
\ea{
R^{\rm IN} =
\max_{\de \in [0,1]} \f 12 \log \lb 1 + \f {\de P}{1+c^2(1+\mu_A^2)+\deb P}\rb + \f 1 {2M} \log \lb 1 +\deb P  \rb  - 1,
}
and  the optimization over $\de$ yields
\ea{
R^{\rm IN} = \lcb \p{
 \f 12 \log \lb 1 + \f{P} {1+c^2(1+\mu_A^2)} \rb & \f {M-1} M > \f 1 M c^2(1+\mu_A^2) \\
%
\f 12 \log (P+c^2 (1+\mu_A^2) +1)   &  \f {M-1} M \leq \f 1 M c^2 (1+\mu_A^2) \leq \f {M-1} M (P+1) \\
\quad - \f {M-1}{2 M } \log \lb c^2 (1+\mu_A^2)  \rb -\Gt  \\
 \f {1} {2M} \log(1+P) - \Gt    & \f 1 M c^2 (1+\mu_A^2) > \f {M-1} M (P+1)
}\rnone
\label{eq:inner bound strong}
}

\bigskip
\noindent
$\bullet$ {\bf Gap between inner and outer bound}
\medskip

The gap between inner and outer bound is obtained by comparing the expressions in \eqref{eq:outer bound M} and the expression in \eqref{eq:inner bound strong}.

\subsection{Proof of Th. \ref{thm:approximate capacity continuous}}
\label{app:approximate capacity continuous}

Similarly to the proof of Th. \ref{thm:approximate capacity discrete mass>1/2} in App. \ref{app:approximate capacity discrete} when deriving an
outer bound to capacity.
%
\eas{
 N(R- \ep)  & \leq I(Y^N;W|A^N) \\
& \leq \f N2 \Ebb_A \lsb \log2 \pi e(P+a^2 c^2  +1 )\rsb - \f N 2 H(Y^N|W,A^N) \\
& \leq \f N2 \log2 \pi e(P+(1+\mu_A)c^2  +1 )- \f N 2 \int_{a^N \in \Acal^N} P(a^N) H(Y^N|W,A^N=a^N) \diff a^N,
\label{eq:fano V1 1}
}{\label{eq:fano V1}}
where
\ean{
H(Y^N|A^N,W)
& = \int_{I^N}P(a^N)  H(X^N +c a S^N+Z^N |W) \diff a^N  \\
& \quad \quad +      \int_{\Rbb^N \setminus I^N} P(a^N)   H(X^N +c a S^N+Z^N|W) \diff a^N.
}
Given the condition in \eqref{eq:condition I set} and since
\ea{
\f N 2  \log( 2 \pi e ) \leq H(X^N +c a S^N+Z^N|W) \leq \f N 2 \log( P + c^2 +1)+1,
}
and $I^N$ is a closed interval, we can apply the mean value theorem   and conclude that
\ea{
\int_{I^N}P(a^N)   H(X^N +c a^N S^N+Z^N|W, A^N=a^N) \diff a^N  = P_A(I^N)  H(X^N +c a'^N S^N+Z^N|W),
}
for some $a'^N \in I^N$.
Note that this holds even if the distribution $P_{X^N,S^N}$ has some discrete points because of the convolution with the distribution of $Z^N$.

We can now write
\ean{
&\int_{I^N}P(a^N)   H(X^N +c a S^N+Z^N|W) \diff a^N \\
& \quad \quad +  \int_{\Rbb^N \setminus I^N} P(a^N)   H(X^N +c a S^N+Z^N|W) \diff a^N  \\
& =  (P_A(I)-(1-P_A(I)))^N H(X^N +c a'^N S^N+Z^N|W) \\
& \quad \quad +  \int_{\Rbb \setminus I} P_A(a)   \lb H(X^N +c a S^N+Z^N|W)+H(X^N +c a'^N S^N+Z^N|W)  \rb \diff a^N  \\
& \geq (P_A(I)-(1-P_A(I)))^N H(X^N +c a'^N S^N+Z^N|X^N,S^N) \\
& \quad \quad +\int_{\Rbb^N \setminus I^N} P(a^N)   \lb H(X^N +c a^N S^N+Z^N,X^N +c a'^N S^N+Z^N|W)  \rb \diff a^N  \\
& \geq N \f{P_A(I)-(1-P_A(I))^N} 2 \log(2\pi e)   \\
& \quad \quad +\int_{\Rbb^N \setminus I^N} P(a^N)   \lb H(c (a^N-a'^N)S^N+Z^N,X^N +c a'^N S^N+Z^N|W)  \rb \diff a^N  \\
%
%
& =N \f{P_A(I)-(1-P_A(I))} 2 \log(2\pi e) \\
& \quad \quad + \int_{\Rbb \setminus I} P(a^N)   \lb \f 12 \log 2 \pi e \lb c^2 (a^N-a'^N)^2 +2 \rb   \rnone \\
& \quad \quad  \lnone+ H(Z^N|c (a^N-a'^N)S^N+Z^N,S^N,X^N)\rb \diff a^N \\
& \geq N \f{P_A(I)} 2 \log(2\pi e)+  \f {P_A(\Rbb \setminus I)} 2 \log  2 \pi e (1+\mu_A^2) c^2  \\
& \quad \quad + N \int_{\Rbb \setminus I} \f {P_A(a)} 2 \log \lb \f { (a-\ao)^2 }{1+\mu_A^2}\rb \diff a \\
& \geq N \f{P_A(I)} 2 \log(2\pi e)+  \f {P_A(\Rbb \setminus I)} 2 \log  2 \pi e (1+\mu_A^2) c^2 + \Gh.
}
This yields the same outer bound as \eqref{eq:outer before optimization} but with an updated expression for $G$.
As for Thm. \ref{thm:approximate capacity discrete mass>1/2} we can optimize the expression in $c$ and obtain the same outer bound.

\end{document}